\documentstyle[aps,preprint,tighten]{revtex}
\newcommand{\lsim}{\mathrel{\mathop{\kern 0pt \rlap
  {\raise.2ex\hbox{$<$}}}
  \lower.9ex\hbox{\kern-.190em $\sim$}}}
\newcommand{\gsim}{\mathrel{\mathop{\kern 0pt \rlap
  {\raise.2ex\hbox{$>$}}}
  \lower.9ex\hbox{\kern-.190em $\sim$}}}

\begin{document}

\preprint{
\begin{tabular}{r}
DFTT 61/97
\end{tabular}
}

\title{Extending a previous analysis on a possible modulation effect
                 in WIMP direct search}

\author{
\bf A. Bottino$^{\mbox{a}}$
\footnote{E--mail: bottino@to.infn.it, donato@to.infn.it,
fornengo@to.infn.it, scopel@posta.unizar.es},
F. Donato$^{\mbox{a}}$, N. Fornengo$^{\mbox{a}}$, 
S. Scopel$^{\mbox{b}}$\footnote{INFN Post--doctoral Fellow}
\vspace{6mm}
}

\address{
\begin{tabular}{c}
$^{\mbox{a}}$
Dipartimento di Fisica Teorica, Universit\`a di Torino and \\
INFN, Sezione di Torino, Via P. Giuria 1, 10125 Torino, Italy
\\
$^{\mbox{b}}$ Instituto de F\'\i sica Nuclear y Altas Energ\'\i as, \\
Facultad del Ciencias, Universidad de Zaragoza, \\
Plaza de San Francisco s/n, 50009 Zaragoza, Spain
\end{tabular}
}
\date{October 9, 1997}
\maketitle

\begin{abstract}
In this  note we present an extension of our previous analysis of some 
preliminary experimental results of the DAMA/NaI Collaboration which might be
indicative of a yearly modulation effect. Here we present a direct way for 
obtaining from the experimental data the
relevant cosmological implications for relic neutralinos.
We find that some of the configurations singled out by the DAMA/NaI 
results would have cosmological properties compatible with a neutralino 
as a dominant component of cold dark matter (on the average in the 
Universe and in our galactic halo).
\end{abstract}  

Some recent results of the DAMA/NaI Collaboration on direct WIMP 
detection \cite{dama} seem to indicate that a yearly modulation 
effect might be present. As stressed in Ref. \cite{dama}, a much larger 
statistics  is necessary, before any claim of a real effect can be 
made; and, actually, a further collection of data is under way at the 
large--mass, low--background NaI(Tl) detector at the Gran Sasso 
Laboratory \cite{dama}. Meanwhile, the preliminary results are so 
intriguing to prompt an analysis of their possible implications  
for particle physics and cosmological properties.
 
   In Ref. \cite{noi}, we have analyzed some possible consequences for 
neutralino properties implied by the results of Ref. \cite{dama}, 
under the hypothesis that these are indicative of a yearly 
modulation effect due to relic neutralinos. We have shown that a number of 
neutralino configurations, compatible with all current experimental 
constraints, might produce a modulation effect at the level of the one 
possibly detected by the DAMA/NaI Group. Some appealing features of these 
configurations (or at least of a subset of these) are: 
i) sizeable cosmological density $\Omega$, ii) detectability of indirect 
signals at neutrino telescopes, iii) detectability 
of some related properties at accelerators. 

     In our previous paper \cite{noi} 
the comparison of theoretical evaluations with experimental data 
was performed in terms of a plot in the plane 
$\xi \sigma^{(\rm nucleon)}_{\rm scalar}$ vs. $m_\chi$, 
where $m_\chi$ is the neutralino mass, 
$\sigma^{(\rm nucleon)}_{\rm scalar}$ is the  scalar elastic
cross section off nucleon and $\xi = \rho_\chi / \rho_l$ 
is the fractional amount of local 
neutralino dark matter density $\rho_\chi$ with respect to the total local 
dark matter density $\rho_l$. The results are reported in Fig. 1
(which is taken from Ref. \cite{noi}).
The closed contour line denotes the region ($R_m$) singled out
at 90\% C.L. by the data of Ref. \cite{dama}.
The open curve denotes the 90\% C.L. upper bound
of $\xi \sigma^{(\rm nucleon)}_{\rm scalar}$, as obtained from the 
total counting rates of Ref. \cite{dama1}. 
In extracting both the open and the closed contour lines
from the experimental data, the values of some 
astrophysical parameters relevant
for the event rates at the detector have to be chosen.
These parameters are: the root mean square velocity 
$v_{\rm rms}$ of the neutralino 
Maxwellian velocity distribution in the halo,
the neutralino escape velocity $v_{\rm esc}$ in the halo, 
the velocity $v_\odot$ of the Sun around the galactic centre.
The values adopted in Fig. 1 
refer to the median values of these parameters in their
experimentally allowed ranges \cite{limiti}, namely: 
$v_{\rm rms} = 270$ Km s$^{-1}$, $v_{\rm esc} = 650$ Km s$^{-1}$,
$v_\odot = 232$ Km s$^{-1}$.
We remark that the values of 
$\xi \sigma^{(\rm nucleon)}_{\rm scalar}$,
extracted from experimental data, are not very sensitive to the 
specific values adopted for the various velocities within
their physical ranges.
The values of $\xi \sigma^{(\rm nucleon)}_{\rm scalar}$
plotted in Fig. 1 are normalized to the value $\rho_l=0.5$ GeV cm$^{-3}$.

Together with the experimental results, 
Fig. 1 shows our evaluations of the
quantity $\xi \sigma^{(\rm nucleon)}_{\rm scalar}$ 
inside the Minimal Supersymmetric
Standard Model. For a brief discussion of the model
we are employing, the reader is addressed to Ref. \cite{noi}.
Here we only recall that the supersymmetric parameter
space, considered here and in Ref. \cite{noi},
consists of six independent parameters: 
$M_2, \mu, \tan\beta, m_A, m_0, A$. 
We vary these parameters in the following ranges: 
$10\;\mbox{GeV} \leq M_2 \leq  500\;\mbox{GeV}$
(21 steps over a linear grid); 
$10\;\mbox{GeV} \leq |\mu| \leq  500\;\mbox{GeV}$
(21 steps, linear grid); 
$65\;\mbox{GeV} \leq m_A \leq  500\;\mbox{GeV}$
(15 steps, logarithmic grid); 
$100\;\mbox{GeV} \leq m_0 \leq  500\;\mbox{GeV}$
(5 steps, linear grid); 
$-3 \leq {\rm A} \leq +3$
(5 steps, linear grid); 
$1.01 \leq \tan \beta \leq 50$
(15 steps, logarithmic grid).

The supersymmetric parameter space is constrained by
the experimental limits obtained from accelerators on
supersymmetric and Higgs searches and from the 
$b \rightarrow s + \gamma$ radiative decay. For a complete
set of references, see Ref. \cite{noi}.
In addition to the experimental limits, the
parameter space is further constrained by the
requirement that the neutralino is the lightest supersymmetric particle,
since the neutralino is the dark matter candidate under
investigation in the present analysis.
Finally, the regions of the parameter space where
the neutralino relic abundance exceeds the cosmological bound, i.e. 
$\Omega_{\chi}h^2 > 1$, are also excluded.

Our analysis in Ref. \cite{noi} was performed employing, for the 
evaluation of $\xi$, the usual rescaling recipe 
\cite{gaisser} which compares
the amount of neutralino relic density $\Omega_\chi h^2$ to the
minimal value for the total amount of dark matter in the Universe
compatible with observational data and with large--scale
structure calculations. The recipe works as follows:
when $\Omega_\chi h^2$ is larger than 
$(\Omega h^2)_{\rm min}$ we simply put $\xi=1$;
when $\Omega_\chi h^2$ turns out to be less than $(\Omega h^2)_{\rm min}$, 
and then the neutralino may only provide a fractional contribution
${\Omega_\chi h^2 / (\Omega h^2)_{\rm min}}$
to $\Omega h^2$, we take $\xi = {\Omega_\chi h^2 / (\Omega h^2)_{\rm min}}$.

Besides the empirical character of the rescaling procedure, it has 
to be noted that, as was stressed in Ref. \cite{noi}, the value of 
$(\Omega h^2)_{\rm min}$ is rather arbitrary, possibly 
in the range $0.03 \lsim (\Omega h^2)_{\rm min} \lsim 0.3$. 
To be definite, in Ref. \cite{noi} we reported our results for the 
representative value $(\Omega h^2)_{\rm min} = 0.03$, and this
is the value used in the scatter plot of Fig. 1.

In this note we present an extension of our previous 
investigation, which is meant to 
obtain from the experimental data the relevant 
cosmological implications for relic neutralinos, 
in the most direct way without any use of 
rescaling for the neutralino local density. 

The procedure we adopt is the following:
1) We evaluate $\sigma^{(\rm nucleon)}_{\rm scalar}$ and 
$\Omega_\chi h^2$ by varying the supersymmetric 
parameters over the previously defined grid.
2) For any value of 
$[\rho_\chi \sigma^{(\rm nucleon)}_{\rm scalar}]_{\rm expt}=\rho_l [\xi
\sigma^{(\rm nucleon)}_{\rm scalar}]_{\rm expt}$ 
compatible with 
the experimental region $R_m$ we calculate $\rho_\chi$ as given by 
$\rho_\chi = [\rho_\chi \sigma^{(\rm nucleon)}_{\rm scalar}]_{\rm expt} / 
\sigma^{(\rm nucleon)}_{\rm scalar}$
and restrict the values of $m_\chi$ to stay inside the region $R_m$.
3) The results are displayed in a scatter plot in the plane 
$\rho_\chi$ vs. $\Omega_\chi h^2$.

     Examples of our results are given in Figs. 2--4  for a few 
experimentally allowed values of $\xi \sigma^{(\rm nucleon)}_{\rm scalar}$,
taken from Fig. 1 (remember that the normalization value $\rho_l=0.5$  GeV
cm$^{-3}$ is used in this figure).  
Figs. 2,3,4 refer to
$[\xi \sigma^{(\rm nucleon)}_{\rm scalar}]_{\rm expt} = 6 \cdot 10^{-9}$ nbarn, 
$2 \cdot 10^{-9}$ nbarn and $1 \cdot 10^{-9}$ nbarn, respectively.
The two horizontal lines denote the physical range  
0.1 GeV cm$^{-3}$  $< \rho_l <$ 0.7 GeV cm$^{-3}$ for the local density 
of non--baryonic dark matter. This (rather generous) 
range has been established by taking 
into account possible effects on 
$\rho_l$ due to a flattened dark halo \cite{turner1}
and a recent derivation \cite{turner} of the contribution 
of baryons to the local dark matter 
from microlensing data. The solid 
vertical lines  denote the physical band for $\Omega h^2$:
$0.03 \lsim \Omega h^2 \lsim 1$, and the two 
dashed lines give the favored band for the cold dark matter 
contribution to $\Omega$: 
$0.1 < (\Omega h^2)_{\rm CDM} < 0.3$ \cite{bere}. The two tilted 
dot--dashed lines denote the band where linear rescaling procedure for the 
local density is usually applied. The upper dot--dashed line refers to 
a rescaling with $(\Omega h^2)_{\rm min} = 0.03$, the lower one to 
the value $(\Omega h^2)_{\rm min} = 0.3$.
     With the aid of this kind of plot we can classify the supersymmetric 
configurations belonging to region $R_m$ into various categories.
Configurations whose representative points fall above the maximum 
value $\rho_\chi = 0.7$ GeV cm$^{-3}$ 
have to be excluded (we remind that those providing 
an $\Omega_\chi h^2 > 1$ are already disregarded from the very beginning).  
Among the allowed configurations, those falling 
in the region inside 
both the  horizontal and solid vertical lines 
(this region is called $A$ hereafter) are very 
appealing, since they would represent situations where the neutralino 
could have the role of a dominant cold dark matter component; even more so,
if the representative points fall 
in the subregion ($B$) inside the vertical 
band delimited by dashed lines. Configurations which fall inside 
the band delimited by the tilted dot--dashed lines denotes situations 
where the neutralino can only provide a fraction of the cold dark 
matter both at the level of local density and at the level of the 
average $\Omega$. Configurations above the upper dot--dashed line and below 
the upper solid horizontal line would imply an unlikely special 
clustering of neutralinos in our halo as compared to their average 
distribution in the Universe.

It is worth noticing a few important properties of 
the scatter plots shown in Figs. 2--4:

\begin{itemize}

\item [1)]
The scatter plots display a correlation between $\rho_\chi$ and 
$\Omega_\chi h^2$. This feature is expected on the basis of the 
following properties: i) $\Omega_\chi h^2$ is roughly inversely 
proportional to the neutralino pair annihilation cross section, 
ii) at fixed $[\xi \sigma^{(\rm nucleon)}_{\rm scalar}]_{\rm expt}$, 
$\rho_\chi$ is inversely proportional to 
$\sigma^{(\rm nucleon)}_{\rm scalar}$, 
iii) the annihilation cross section and 
$\sigma^{(\rm nucleon)}_{\rm scalar}$ are usually 
correlated functions (i.e., they are both increasing or 
decreasing functions of the supersymmetric parameters). 

\item [2)]  
The regions $A$ and $B$ become more populated 
in configurations as the experimental value 
$[\xi \sigma^{(\rm nucleon)}_{\rm scalar}]_{\rm expt}$
decreases (i.e., if one proceeds from the case shown in Fig. 2 
to the one in Fig. 4). This feature follows from the fact that 
$\sigma^{(\rm nucleon)}_{\rm scalar}$ is 
bounded from above by accelerator limits (mainly because of lower 
bounds on Higgs masses); this implies for $\rho_\chi$ 
a lower bound, which however is less stringent at lower values of 
$[\xi \sigma^{(\rm nucleon)}_{\rm scalar}]_{\rm expt}$.

\item [3)] 
It is remarkable that, as a consequence of 
point 2), some neutralino configurations populate the most 
interesting region, i.e. region $B$. This occurs for 
$[\xi \sigma^{(\rm nucleon)}_{\rm scalar}]_{\rm expt} 
\lsim 6 \cdot 10^{-9}$ nbarn.

\end{itemize}

In what follows we concentrate on configurations falling in 
region $A$ or in the subregion $B$.
For definiteness, we choose the representative value $[\xi \sigma^{(\rm
nucleon)}_{\rm scalar}]_{\rm expt} = 2 \cdot 10^{-9}$ nbarn. 
First, we check whether some of these neutralino 
compositions are excluded by indirect detection of WIMPs, by 
measurements of up--going muon fluxes 
due to the neutrino outburst produced by
neutralino pair annihilation in the Earth or 
in the Sun. In Fig. 5 we show  how our calculated 
fluxes from the Earth $\Phi_{\mu}^{\rm Earth}$
compare with the current experimental upper bounds 
\cite{baksan,macro}. We notice that some configurations are 
actually excluded by the present limits on $\Phi_{\mu}^{\rm Earth}$.
From our calculations, it turns out that the present upper bounds
on the up--going muon fluxes from the Sun constrain 
the analyzed configurations only marginally. 

      Finally, we provide in Figs. 6,7 some information about 
important parameters referring  to the configurations of region 
$A$ and of subregion $B$. These figures also show what may be the 
exploration potential for these configurations at LEP2 and Tevatron. 

      In conclusion, we remark that our present analysis shows,  
in a way independent of the rescaling procedure on local DM 
density, that some of the configurations singled out by the DAMA/NaI 
results would have cosmological properties compatible with 
a neutralino as a dominant component of cold dark matter 
(on the average in the Universe and in our galactic halo).

\newpage
\begin{center}
\begin{large}
FIGURE CAPTIONS
\end{large}
\vspace{5mm}\
\end{center}

\begin{itemize}
\item [] 
{\bf Figure 1} -- 
The scalar neutralino--nucleon cross section 
$\sigma^{\rm (nucleon)}_{\rm scalar}$, multiplied by the 
factor $\xi$, is plotted versus the neutralino mass $m_\chi$.
The closed contour delimits the region
singled out at 90\% C.L. by the data of Ref. \cite{dama}.
The open curve denotes the 90\% C.L. upper bound 
obtained from the total
counting rates of Ref.\cite{dama1}.
The scatter plot represents the theoretical prediction
for $\xi\sigma^{\rm (nucleon)}_{\rm scalar}$, calculated within the
MSSM scheme. Only configurations with $\mu>0$ are displayed.
The cosmological and astrophysical parameters are set to the
following values:
$v_{\rm rms} = 270$ Km s$^{-1}$, $v_{\rm esc} = 650$ Km s$^{-1}$,
$v_\odot = 232$ Km s$^{-1}$, $\rho_l = 0.5$ GeV cm$^{-3}$ and
$(\Omega h^2)_{\rm min} = 0.03$.

\item []
{\bf Figure 2} -- 
The neutralino local density $\rho_\chi$, calculated
for $[\xi \sigma^{(\rm nucleon)}_{\rm scalar}]_{\rm expt} =
6 \cdot 10^{-9}$ nbarn, is plotted versus the
neutralino relic abundance $\Omega_\chi h^2$.
For the value of 
$[\xi \sigma^{(\rm nucleon)}_{\rm scalar}]_{\rm expt}$
employed here, the neutralino mass is restricted to
the range 30 GeV $\lsim m_\chi \lsim 155$ GeV, as
obtained from the closed contour in Fig. 1. 
The two horizontal lines denote the physical range  
for the local density of non--baryonic dark matter.
The two solid vertical lines denote the physical band for 
$\Omega h^2$, and the two dashed lines give the preferred band 
for the cold dark matter contribution to $\Omega$. The two
tilted dot--dashed lines denote the band where linear rescaling 
procedure for the local density is usually applied. 
The region inside both the two horizontal lines and the
two vertical solid (dashed) lines is defined as region $A$
(region $B$).

\item []
{\bf Figure 3} -- 
The neutralino local density $\rho_\chi$, calculated
for $[\xi \sigma^{(\rm nucleon)}_{\rm scalar}]_{\rm expt} =
2 \cdot 10^{-9}$ nbarn, is plotted versus the
neutralino relic abundance $\Omega_\chi h^2$.
The neutralino mass is restricted to
the range 30 GeV $\lsim m_\chi \lsim 115$ GeV, as
obtained from the closed contour in Fig. 1. 
The solid, dashed and dot--dashed lines,
as well as region $A$ and region $B$,
are defined as in Fig. 2.

\item []
{\bf Figure 4} -- 
The neutralino local density $\rho_\chi$, calculated
for $[\xi \sigma^{(\rm nucleon)}_{\rm scalar}]_{\rm expt} =
1 \cdot 10^{-9}$ nbarn, is plotted versus the
neutralino relic abundance $\Omega_\chi h^2$.
The neutralino mass is restricted to
the range 40 GeV $\lsim m_\chi \lsim 85$ GeV, as
obtained from the closed contour in Fig. 1. 
The solid, dashed and dot--dashed lines,
as well as region $A$ and region $B$,
are defined as in Fig. 2.

\item []
{\bf Figure 5} -- 
Flux of up--going muons $\Phi_{\mu}^{\rm Earth}$
as a function of the neutralino mass $m_{\chi}$, calculated
for the neutralino configurations belonging
to region $A$ (dots) and to region $B$ (crosses)
of Fig. 3 (therefore referring to
$[\xi \sigma^{(\rm nucleon)}_{\rm scalar}]_{\rm expt} =
2 \cdot 10^{-9}$ nbarn and
40 GeV $\lsim m_\chi \lsim 115$ GeV).
The solid and dashed lines represent the experimental 
90\% C.L. upper bounds of Ref.\cite{baksan} and
Ref. \cite{macro}, respectively.

\item []
{\bf Figure 6} -- 
The configurations compatible with region $A$ (dots)
and region $B$ (crosses) of Fig. 3 are plotted
in the $m_\chi$--$\tan\beta$ plane. 
Configurations which provide a 
flux of up--going muons 
exceeding the experimental upper limits of Ref. \cite{baksan,macro}
are dropped.
The dark region on the left side is 
excluded by current LEP data\cite{lep2}. The region on the left
of the vertical solid line will be accessible to
LEP at $\sqrt{s} = 192$ GeV\cite{yellow}. The region on the left of
the vertical dashed line will be explorable at TeV33\cite{tevatron}.

\item []
{\bf Figure 7} -- 
The configurations compatible with region $A$ (dots)
and region $B$ (crosses) of Fig. 3 are plotted
in the $m_h$--$\tan\beta$ plane. 
Configurations which provide a 
flux of up--going muons 
exceeding the experimental upper limits of Ref. \cite{baksan,macro}
are dropped.
The dark regions are excluded by current
LEP searches \cite{lep2} or by theoretical arguments.
The region on the left of the solid line 
will be accessible to LEP at $\sqrt{s} = 192$ GeV, 
with a luminosity of 150 pb$^{-1}$ per experiment
\cite{yellow}.

\end{itemize}

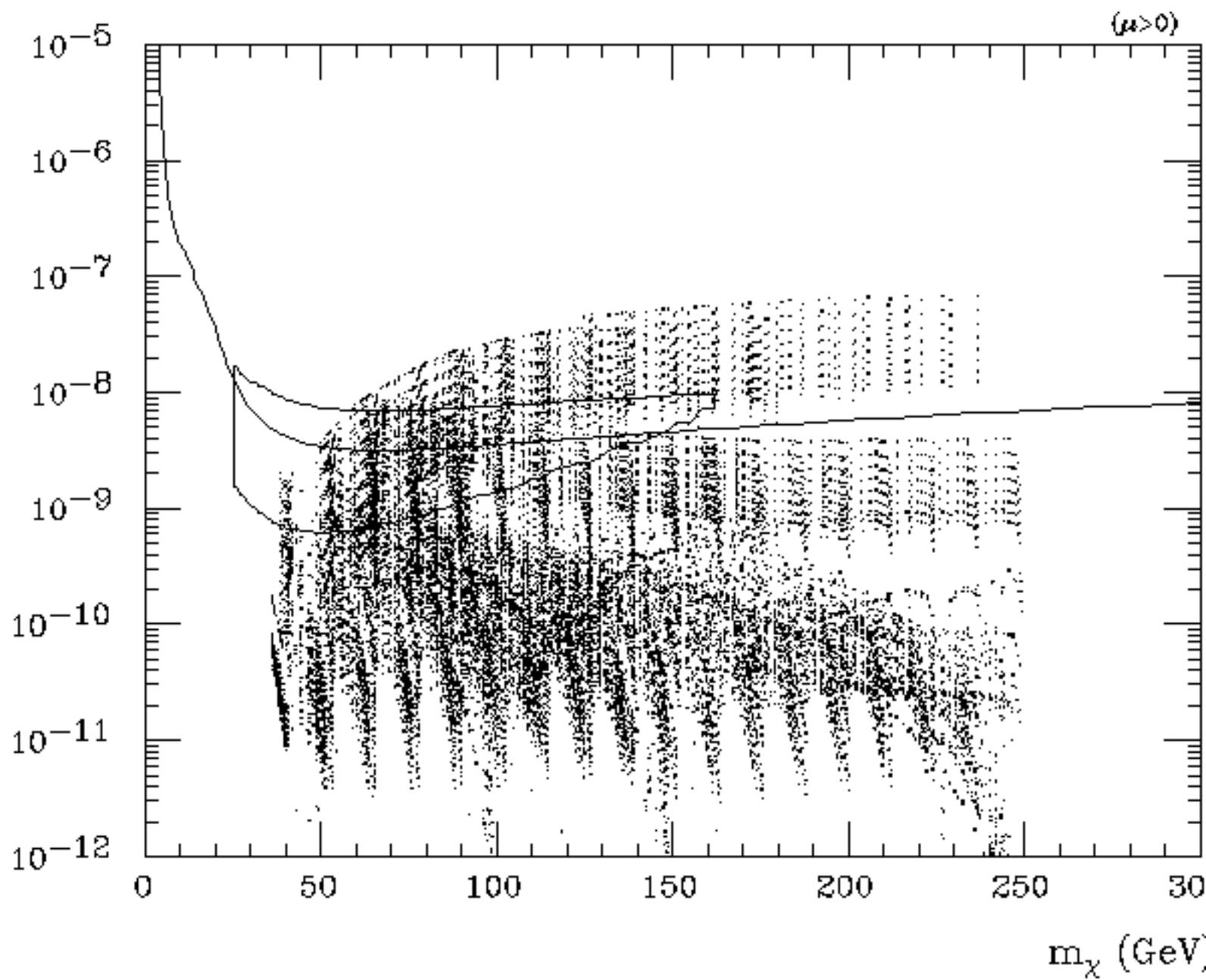
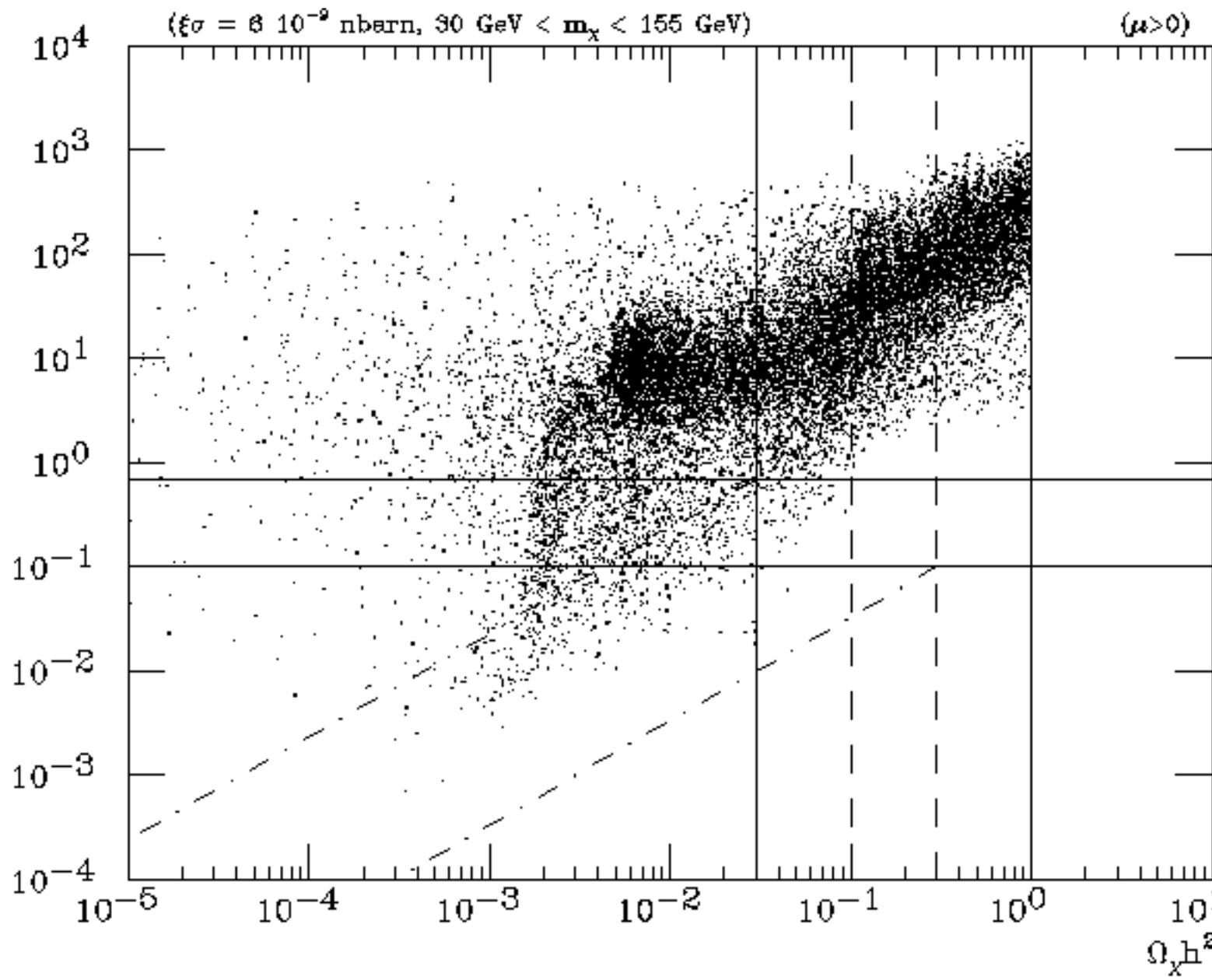
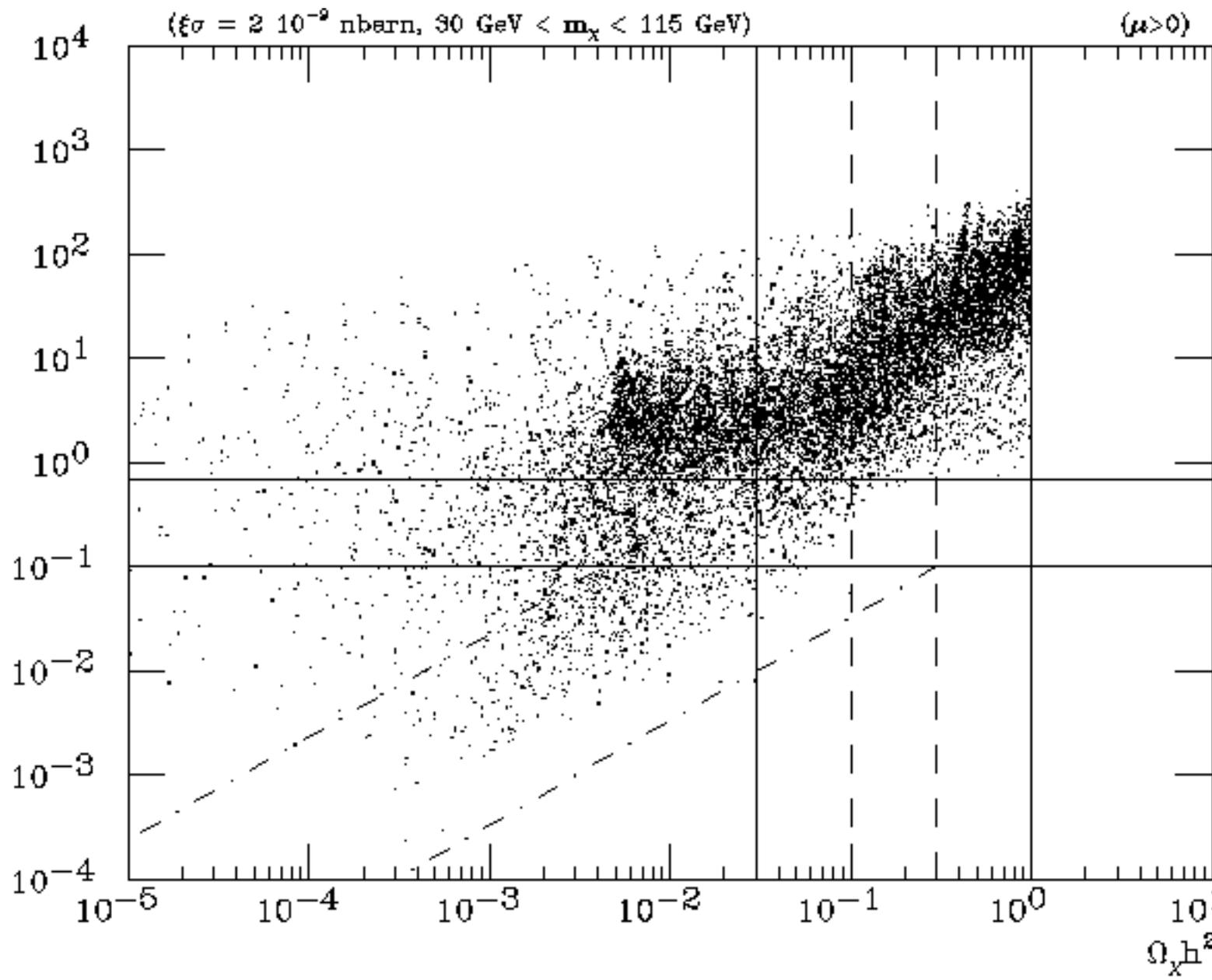
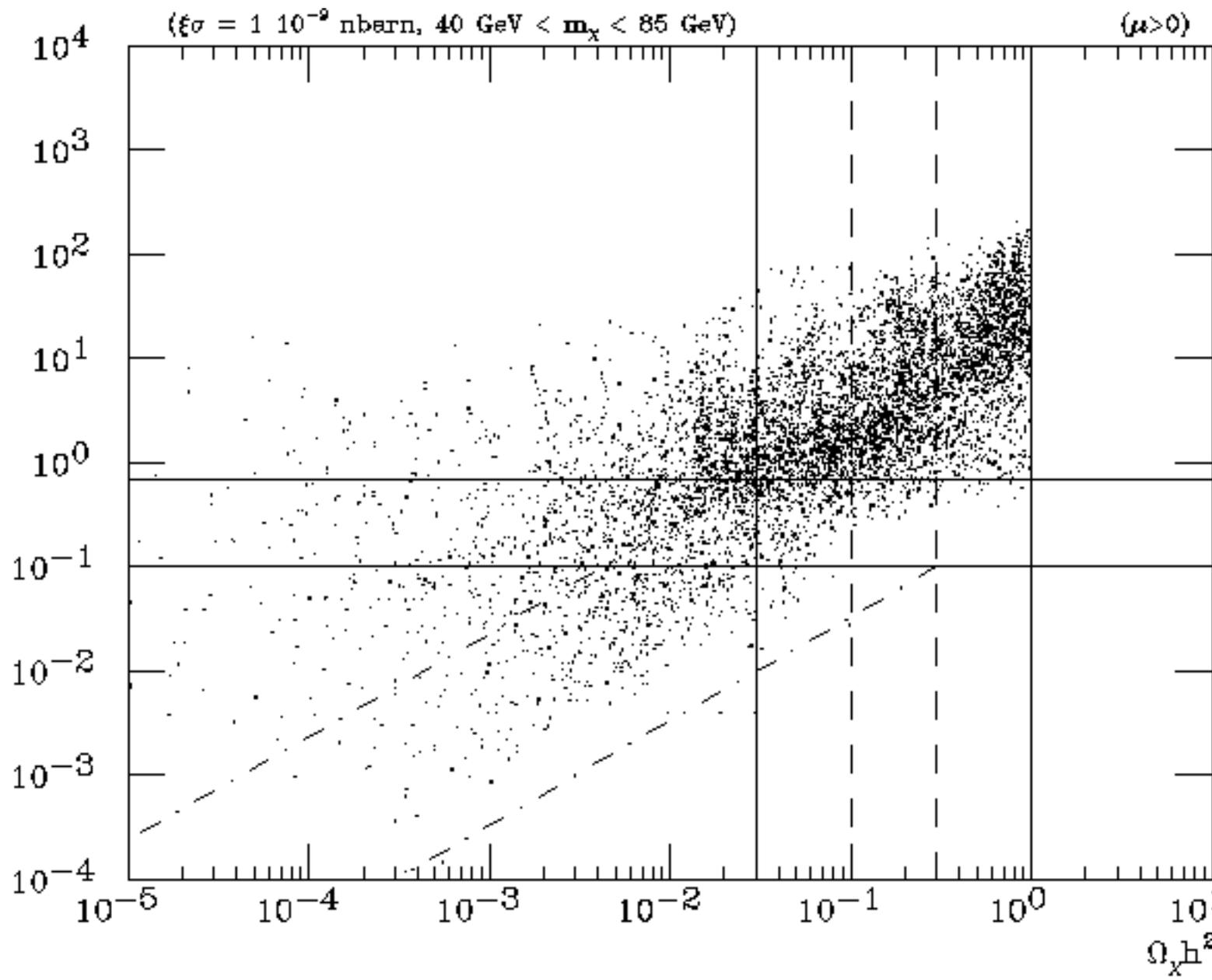
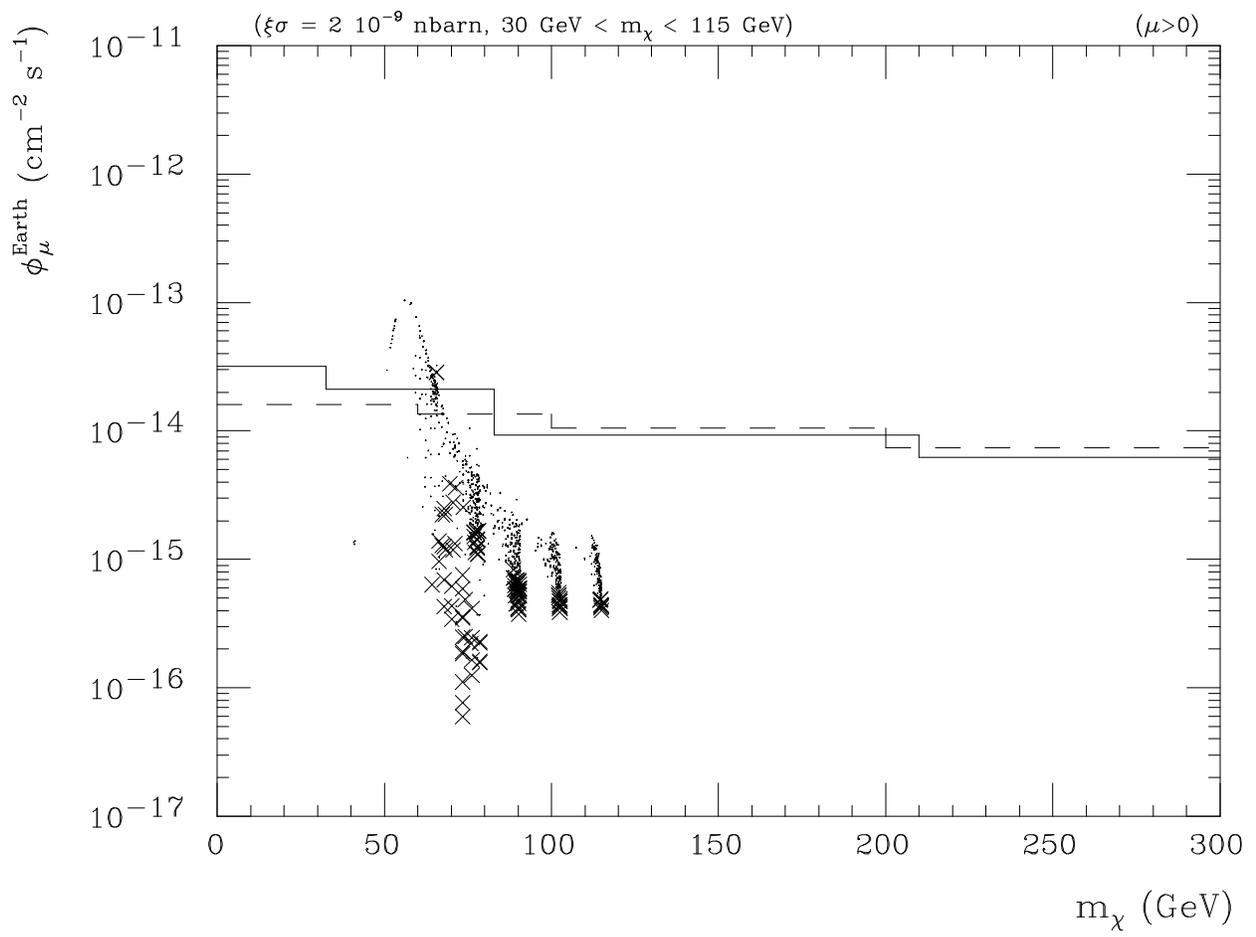
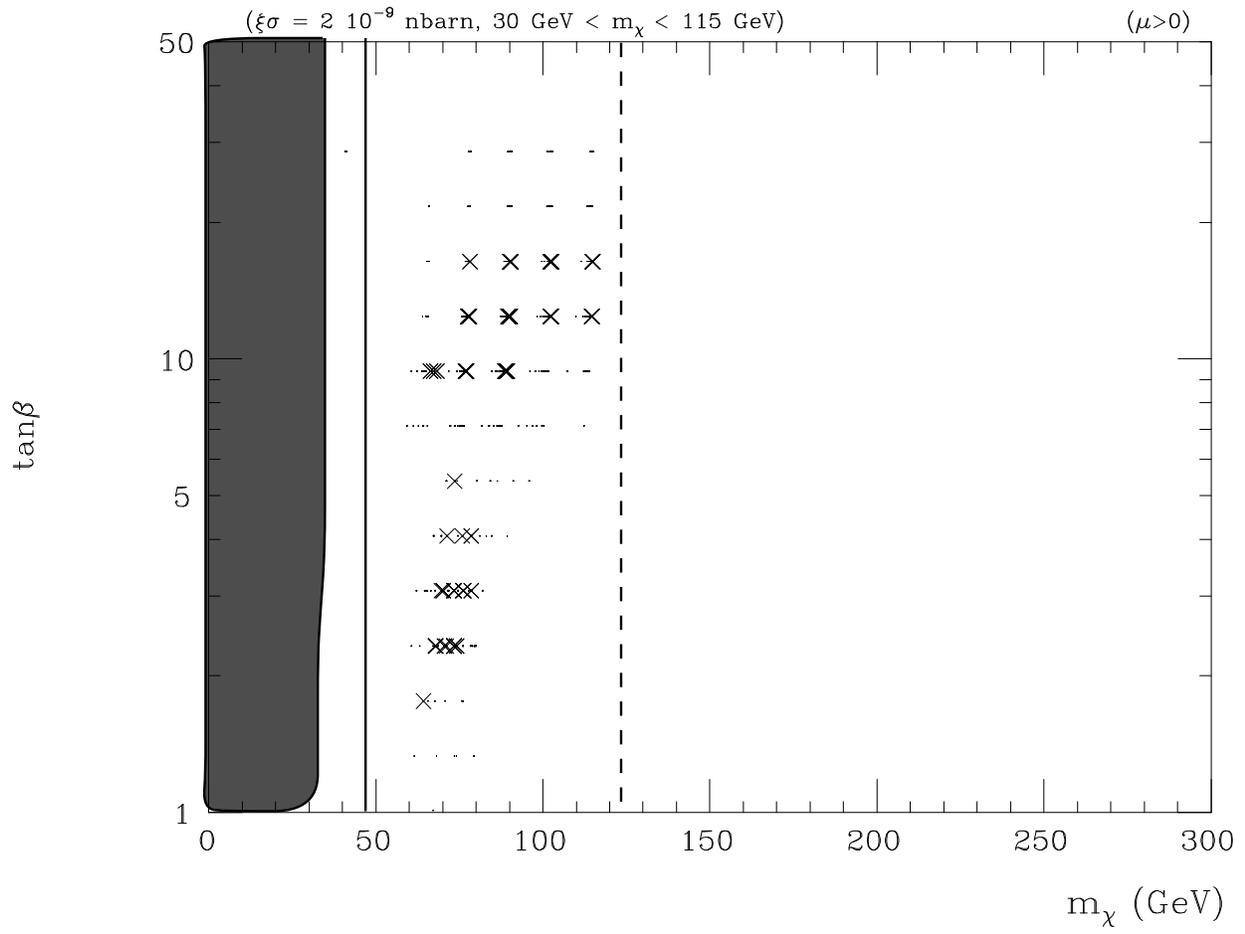
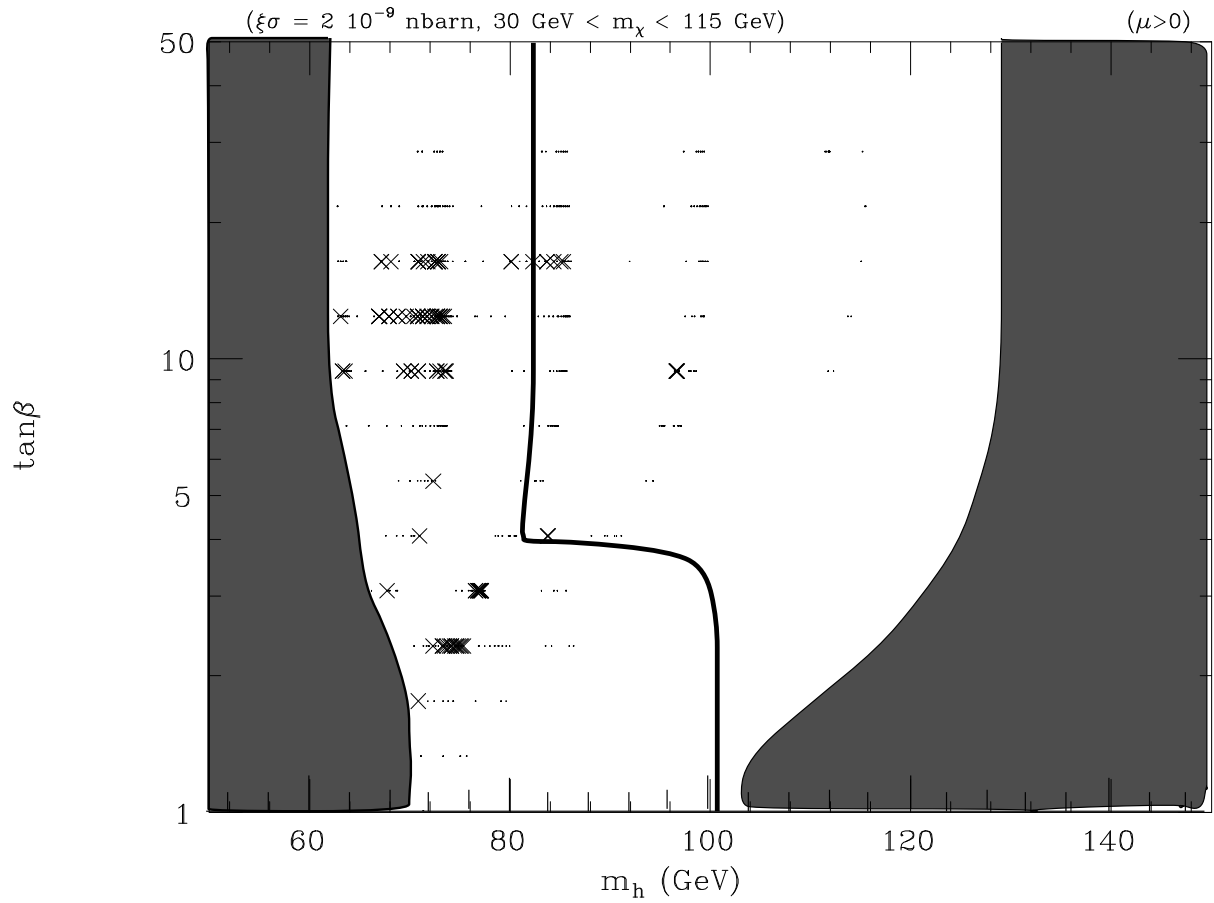
\end{document}